\documentclass{webofc}
\usepackage[varg]{txfonts}   
\usepackage{enumitem}

\newcommand{\GEANT} {{\textsc{Geant}}\xspace}
\newcommand{\GEANTfour} {{\textsc{Geant4}}\xspace}
\newcommand{\GEANTV} {{\textsc{GeantV}}\xspace}
\newcommand{\HEPMC} {{\textsc{HepMC}}\xspace}
\newcommand{\VECGEOM} {{\textsc{VecGeom}}\xspace}
\newcommand{\VECCORE} {{\textsc{VecCore}}\xspace}
\newcommand{\VECMATH} {{\textsc{VecMath}}\xspace}
\newcommand{\CMSSW} {{\textsc{CMSSW}}\xspace}
\newcommand{\EXTERNALWORK} {{\textsc{ExternalWork}}\xspace}
\newcommand{\ROOT} {{\textsc{ROOT}}\xspace}
\newcommand{\TGEO} {{\textsc{TGeo}}\xspace}
\newcommand{\unit}[1]{\ensuremath{\text{\,#1}}\xspace}
\newcommand{\percms}{\ensuremath{\,\text{cm}^\text{$-$2}\,\text{s}^\text{$-$1}}\xspace}
\newcommand{\ttbar}{\ensuremath{{\text{t}\overline{\text{t}}}}\xspace} 

\begin{document}

\title{Integration and Performance of New Technologies in the CMS Simulation}

\author{\firstname{Kevin} \lastname{Pedro}\inst{1}\fnsep\thanks{\email{pedrok@fnal.gov}} (on behalf of the CMS Collaboration)}

\institute{Fermi National Accelerator Laboratory, Batavia, IL, USA}

\abstract{%
The HL-LHC and the corresponding detector upgrades for the CMS experiment will present extreme challenges for the full simulation. In particular, increased precision in models of physics processes may be required for accurate reproduction of particle shower measurements from the upcoming High Granularity Calorimeter. The CPU performance impacts of several proposed physics models will be discussed. There are several ongoing research and development efforts to make efficient use of new computing architectures and high performance computing systems for simulation. The integration of these new R\&D products in the CMS software framework and corresponding CPU performance improvements will be presented.
}

\maketitle

\section{Introduction}\label{sec:intro}

The high luminosity LHC upgrade (HL-LHC) will increase the instantaneous luminosity for proton-proton collisions to 5--$7.5\times10^{34}\percms$,
at least a factor of 5 greater than the design luminosity of the LHC.
This upgrade will produce an order of magnitude more data over the lifetime of the collider,
but it will also produce more radiation, which can damage the detector, and more simultaneous interactions per bunch crossing, or pileup.
The CMS detector will be correspondingly upgraded to cope with these extreme conditions and to maintain its high data quality and strong physics performance.
In particular, the endcap calorimeter system will be replaced with an integrated high granularity calorimeter (HGCal)~\cite{CMS-TDR-019}.

The HL-LHC upgrade and associated detector upgrades have significant implications for the software and computing systems in CMS.
In 2016, at the start of Run 2 of the LHC, the detector simulation step in the CMS software (\CMSSW) chain consumed roughly 40\% of the total CPU usage~\cite{Apostolakis:2018ieg}.
The full detector simulation uses the \GEANTfour software~\cite{Geant4SW,Agostinelli:2002hh,Allison:2016lfl}.
The other steps, including event generation, digitization (electronics simulation), reconstruction, and analysis, consumed 45\% of the total CPU usage,
with the plurality belonging to reconstruction.
The different steps will scale in different ways in the HL-LHC era.
All steps will need to process more simulated events in order to produce samples of appropriate sizes to compare to the large HL-LHC datasets.
However, the CPU usage in the reconstruction step is expected to scale at least linearly with pileup,
and in addition, reconstruction algorithms are expected to become more complex for the future LHC runs.
Therefore, we expect reconstruction to consume a larger fraction of the available CPU in Run 4, compared to Run 2~\cite{Alves:2017she}.

While the detector simulation CPU usage will not scale with pileup, the HGCal detector upgrade presents several challenges, which are described in Section~\ref{sec:chal}.
The detector simulation will need to overcome these challenges while using a smaller fraction of CPU time in Run 4, because of the larger fraction needed for reconstruction.
An overview of the ongoing R\&D to address these needs is presented in Section~\ref{sec:rd}.
The technical details of how these R\&D products are integrated in \CMSSW are given in Section~\ref{sec:integ},
while Section~\ref{sec:results} shows the performance results from the integration.

\section{Challenges}\label{sec:chal}

The HGCal geometry is significantly more complicated than the calorimeters that are currently part of the CMS detector.
HGCal will have approximately 6 million channels~\cite{CMS-TDR-019},
compared to the 91,000 channels in the Run 3 electromagnetic and hadronic calorimeters (ECAL and HCAL)~\cite{Chatrchyan:2008zzk,CMS-TDR-010}.
The increase in complexity is reflected in the simulated geometry for HGCal, which contains ten times more volumes than the present detector geometry.
By itself, this causes the simulation to take 40--60\% longer~\cite{Apostolakis:2018ieg}.

However, the changes needed to simulate the HGCal do not just impact the geometry.
The complexity of the calorimeter enables measurements of particle showers with unprecedented precision.
Therefore, the precision of the simulation must also increase.
\GEANTfour simulates particle interactions using various models with different kinematic or geometric regions of validity;
these models are grouped together to form ``physics lists''.
CMS currently uses a modified physics list called \textsc{ftfp\_bert\_emm}, which uses a simplified multiple scattering model
in most detector regions, except for HCAL and HGCal.
This modification reduces the CPU usage of the simulation by 15\%, compared to the default \textsc{ftfp\_bert} physics list.

\begin{figure}[h]
\centering
\includegraphics[width=0.75\linewidth]{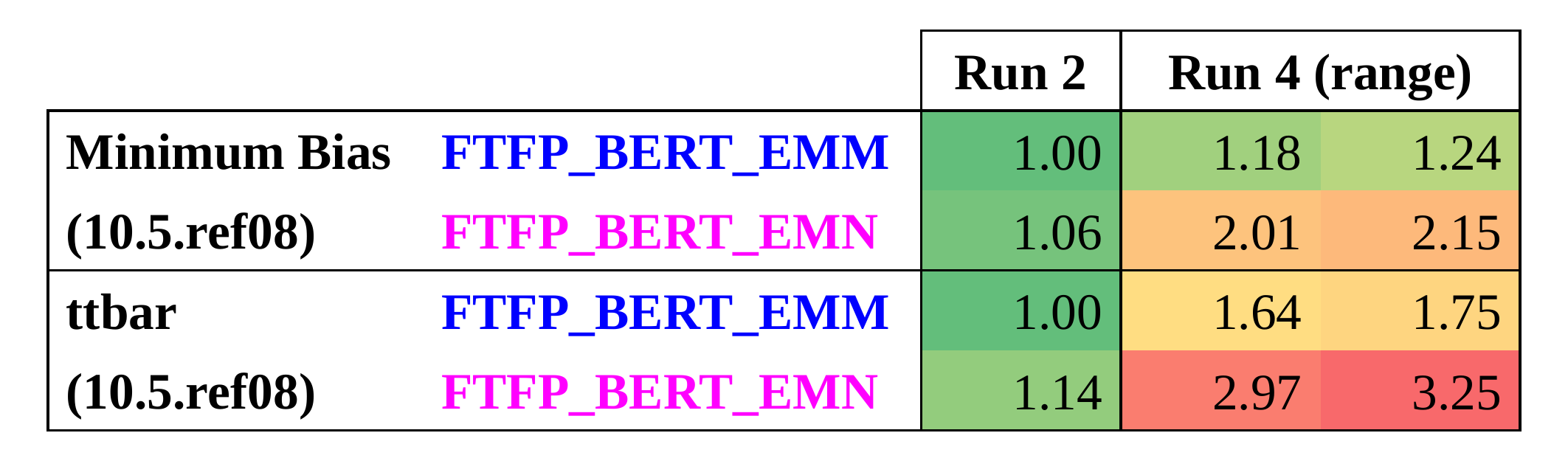}
\caption{Detector simulation CPU usage for two different processes, minimum bias and \ttbar, comparing two different physics lists and the Run 2 and Run 4 CMS geometries.
The value for each entry is normalized to the Run 2 \textsc{ftfp\_bert\_emm} value for the corresponding process.}
\label{fig:phys_list_table}
\end{figure}

A new, prototype physics list \textsc{ftfp\_bert\_emn} is developed to test the impact of increased precision on the simulation.
This list includes the Goudsmit-Saunderson model~\cite{KADRI20093624} for $\text{e}^{+}\text{e}^{-}$ multiple scattering below 100\unit{MeV},
a new angular generator for bremsstrahlung, and a more accurate Compton scattering model~\cite{Allison:2016lfl}.
The CPU usage with the new physics list is compared to the usage with the existing CMS physics list in Fig.~\ref{fig:phys_list_table},
using the latest available version of \GEANTfour and two different processes, minimum bias and \ttbar.
A range of values is given for the Run 4 result based on different upgraded detector geometry configurations.
It can be seen that the detector simulation takes 2--3 times longer when the new, more precise physics list is used with HGCal.
The measurements in this section were performed with recent \CMSSW releases~\cite{CMSSW},
specifically the 9 and 10 series, which have similar performance.

\section{Research and development}\label{sec:rd}

The CMS collaboration constantly tests technical improvements and physics-preserving approximations to improve the CPU performance of the detector simulation.
This ongoing effort has produced impressive results, increasing the performance of the simulation by a factor of 4--6 compared to the default \GEANTfour settings~\cite{Pedro:2019mkq}.
The major contributions to this improvement include the Russian roulette algorithm~\cite{Lange:2015sba}, the use of shower libraries in the forward region,
and the implementation of specific production cuts and physics models for different detector regions.
The inclusion of new R\&D products is also important, such as \VECGEOM~\cite{VecGeomSW,Apostolakis:2015raa}, which brings a 7--14\% improvement.
However, the HL-LHC era will bring significant demands for the simulation, as indicated in Section~\ref{sec:chal}.
New approaches are needed to improve the software to the point that it can run the detailed HGCal detector simulation in a smaller fraction of the experiment's total CPU.

One promising new approach is the \GEANTV vectorized transport engine.
\GEANTV is a prototype that performs the same operations as \GEANTfour---geometry navigation, magnetic field propagation, particle interactions with materials via physics lists---
with a completely rewritten code base.
The initial goal of the \GEANTV R\&D project was to make better usage of modern CPU registers by improving data locality.
It includes track-level parallelism, with similar tracks from multiple events grouped into baskets and processed simultaneously.
This allows the exploitation of single instruction, multiple data (SIMD) vectorization.
The software is built on abstraction libraries that automatically produce efficient code for different CPU instruction sets and even new, heterogenous computing architectures.
These libraries include \VECCORE~\cite{VecCoreSW}, \VECMATH~\cite{VecMathSW}, and the aforementioned \VECGEOM.

\hyphenation{vectorization}
After several years of development and testing, standalone tests of \GEANTV find that it is a factor of $2.0\pm0.5$ faster than similar \GEANTfour configurations~\cite{amadio2020geantv}.
The primary source of this CPU performance improvement is found to be reduced instruction cache misses, while vectorization and other data locality benefits have minor effects.
The new source code uses modern C++ features, producing a smaller compiled library that fits more easily into CPU caches; this facilitates switching between the different operations involved in detector simulation.
Correspondingly, the improvement factor depends strongly on the size of the CPU cache; CPUs with smaller caches see a larger improvement with \GEANTV.
The use of vectorization improves the simulation CPU performance by 15--30\%, depending on the instruction set used,
and with the caveat that the majority of operations in the software have not yet been vectorized.
Considering just the electromagnetic physics operations, a factor of 1.5--4 using AVX2 instructions on Haswell or Skylake chips is achieved via vectorization.
However, the impact of this improvement on the overall simulation time is limited by the presence of numerous other components, some of which are not vectorized.
In addition, the scheduling operations needed to organize the basketization of tracks have significant overhead.
Nevertheless, improving the CPU usage of the detector simulation by even a small integer factor could significantly ameliorate the challenges of the HL-LHC era.

\section{Integration of new technologies}\label{sec:integ}

In this section, the technical details of the effort to integrate \GEANTV in \CMSSW are presented.
This effort has several goals:
\begin{enumerate}[nosep]
\item to engage in co-development between the \GEANTV developers and the experiment software developers, in order to prevent any divergences or incompatibilities;
\item to measure any potential CPU penalties from running \GEANTV in the full experiment software framework, as opposed to standalone tests;
\item to estimate the human cost for the experiment to adapt to new interfaces when considering migration to new, backward-incompatible tools.
\end{enumerate}

The resulting code is publicly available in Ref.~\cite{SimGVCoreSW}.
Physics events can be generated in \CMSSW in the \HEPMC format~\cite{HepMC2SW,Dobbs:2001ck} and converted to the \GEANTV native format.
The CMS detector geometry is built natively in \CMSSW~\cite{Osborne_2014} and passed to the \GEANTV engine using the {\ROOT} \TGEO library~\cite{RootSW,Brun:1997pa}.
The scoring code for ECAL and HCAL, which converts \GEANT step information into simulated detector hits, has been adapted to the \GEANTV interfaces.
These detector hits are produced in the \CMSSW format and therefore are immediately suitable for use in downstream steps, such as digitization.
Currently, a constant magnetic field with $B = 3.8\unit{T}$ is used, along with a physics lists that only includes models of electromagnetic processes;
these limitations reflect what is available in \GEANTV.

\begin{figure}[h]
\centering
\includegraphics[width=0.95\linewidth]{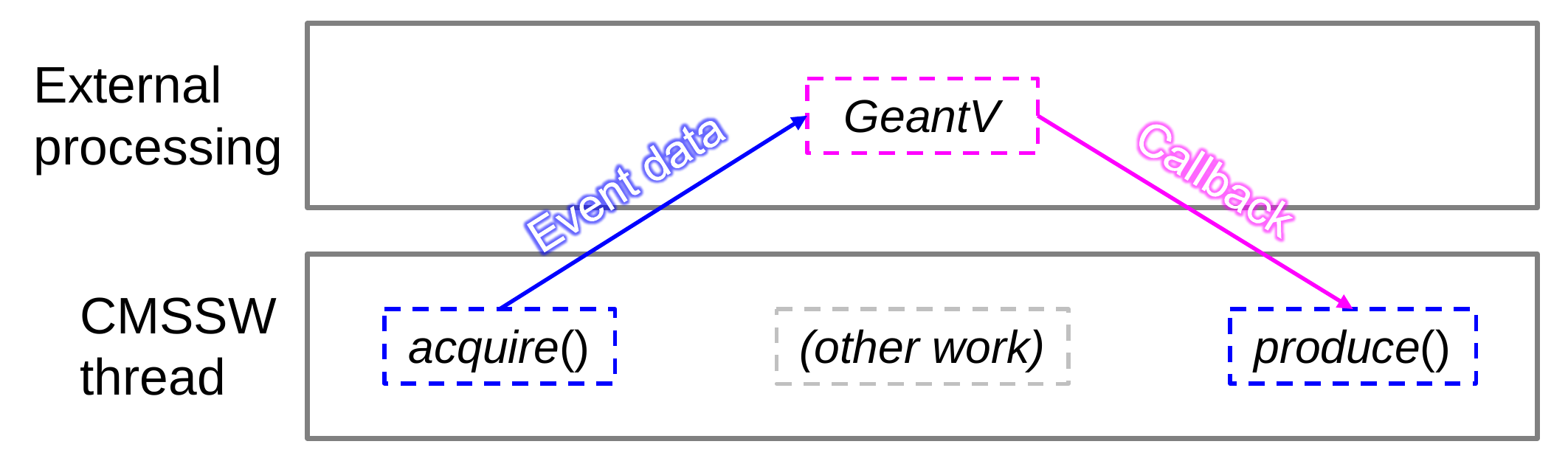}
\caption{A diagram illustrating the asynchronous, non-blocking multithreaded processing enabled by \EXTERNALWORK.}
\label{fig:externalwork}
\end{figure}

\CMSSW uses its \EXTERNALWORK feature~\cite{makortelCHEP2019} to run \GEANTV. \EXTERNALWORK enables asynchronous, non-blocking, task-based multithreaded processing;
this is a key detail of the implementation, depicted in Fig.~\ref{fig:externalwork}. \GEANTV processes multiple events in parallel, mixing the tracks from those events in different threads.
If an event is loaded in one thread, its last track may finish processing in a different thread.
Therefore, sending the generator-level event to be processed in \GEANTV must be a different task
than receiving the simulated hits when \GEANTV is finished processing the event, and such a workflow is enabled by \EXTERNALWORK.

\begin{figure}[h]
\centering
\includegraphics[width=0.95\linewidth]{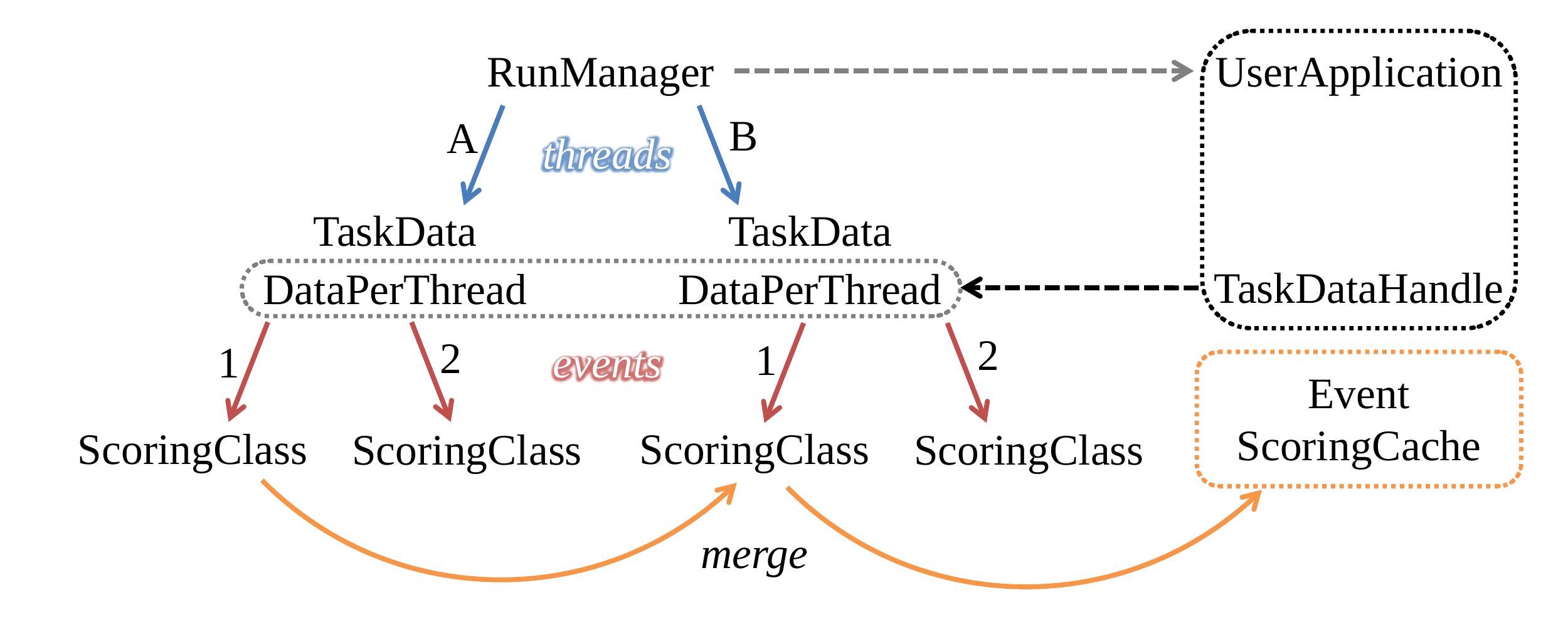}
\caption{A schematic diagram that shows the process of aggregating scoring data in \GEANTV.}
\label{fig:taskdata}
\end{figure}

The mixing of tracks from different events across different threads also complicates the adaptation of the scoring code.
In multithreaded usage of \GEANTfour, each event is processed in a separate thread, so only one instance of the scoring class object per thread is needed.
In \GEANTV, each scoring class must be duplicated per thread and per event stream, in order to avoid significant code changes
that would be needed to support multiple threads modifying the same object simultaneously.
This approach incurs some additional memory overhead from duplicated class members, but this can be minimized
by identifying members that are not modified during event processing and sharing them between instances.
When an event is finished processing, its simulated hits are thus spread across multiple scoring class objects, and must be aggregated.
\GEANTV supports this aggregation using its \textsc{TaskData} interface, which is depicted in Fig.~\ref{fig:taskdata}.
The merged output is copied to a cache attached to the event object, which is also accessible to \CMSSW.

There is an additional challenge in adapting the scoring code to work with \GEANTV.
The total scoring code in \CMSSW comprises roughly 10,000 lines of code, and it relies on \GEANTfour interfaces.
The interfaces in \GEANTV are completely different from \GEANTfour,
but the experiment lacks the personpower to rewrite and validate such a large amount of complicated and delicate code.
To solve this problem, the scoring classes are modified to become class templates, with the template parameter
taken to be a traits class that provides wrappers to unify the disparate interfaces.
The scoring code then only calls accessors from the wrappers, which store pointers to the underlying \GEANTfour or \GEANTV objects.
This avoids any copying, branching, or virtual table calls, minimizing overhead to preserve the computing performance.
Wrappers are written for the run, event, step, and volume classes, providing a path to integration
that preserves the existing scoring code and clearly shows the correspondence between \GEANTfour and \GEANTV operations.
Thus, the exact same scoring code is used for both \GEANTfour and \GEANTV.

\section{Results}\label{sec:results}

\begin{figure}[h]
\centering
\includegraphics[width=0.49\linewidth]{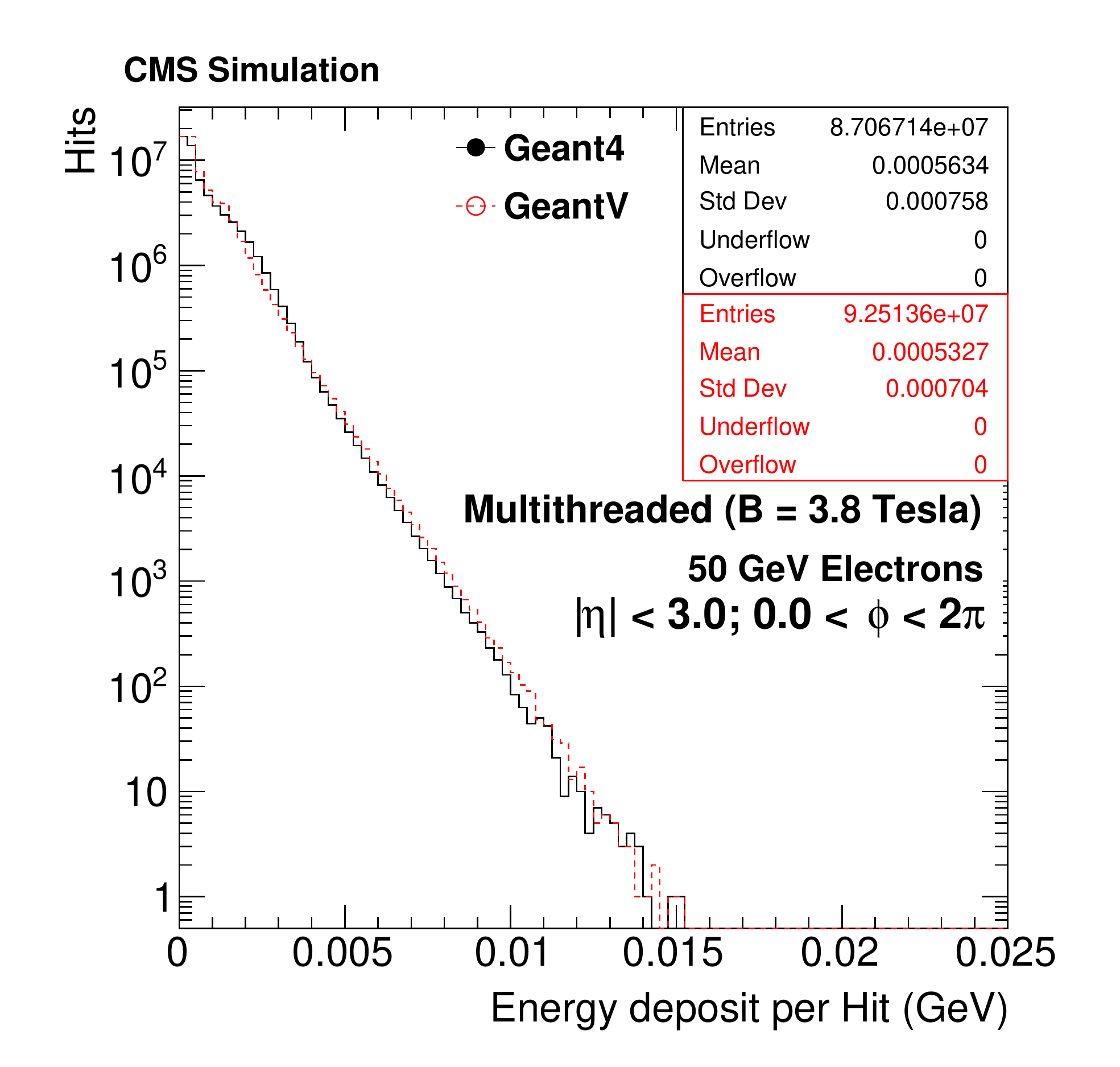}
\includegraphics[width=0.49\linewidth]{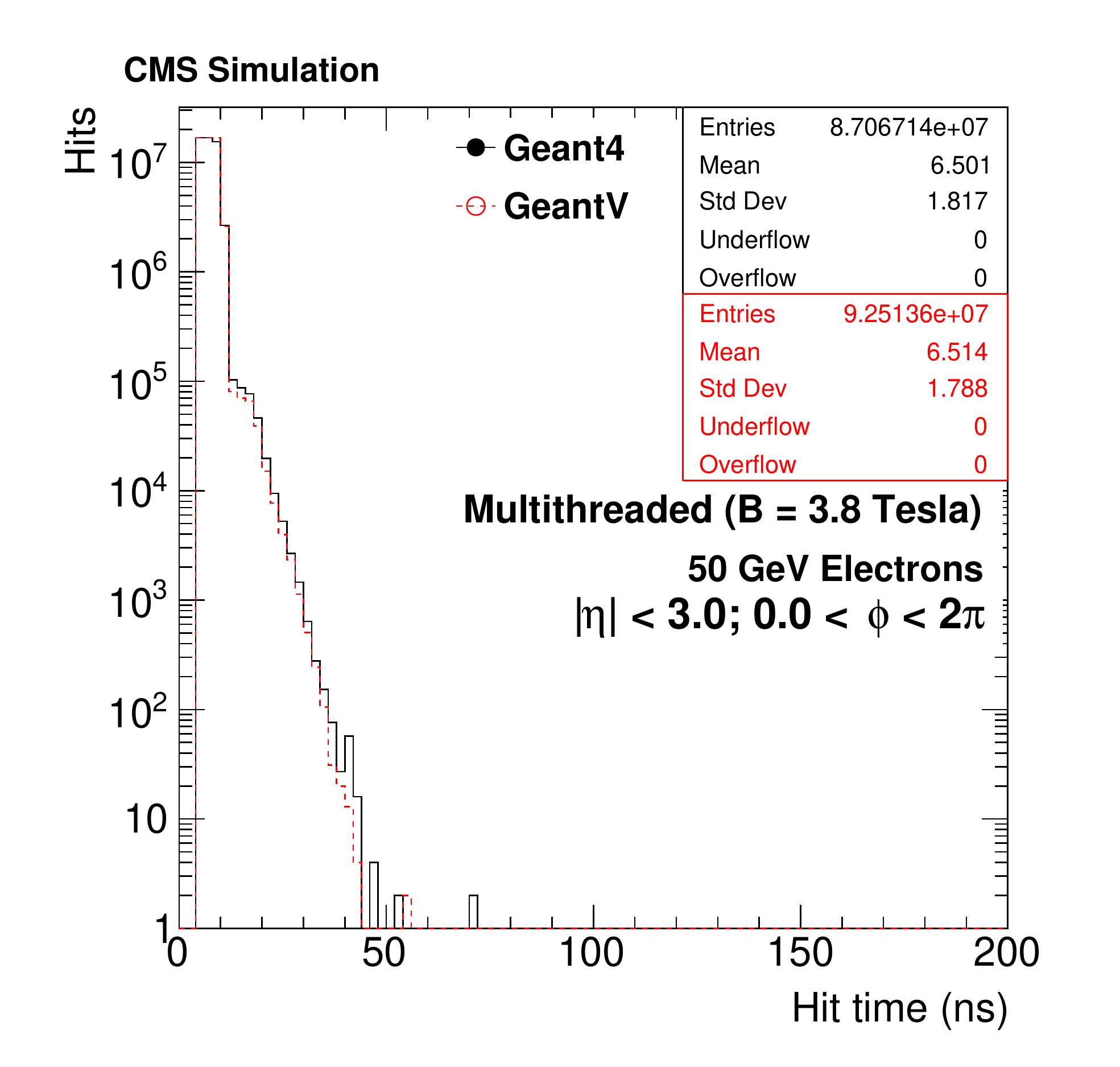}
\caption{The distributions of deposited energy and time for each simulated hit in the ECAL barrel, comparing \GEANTfour and \GEANTV.
The hits originate from a simulation of single electrons with $E=50\unit{GeV}$ and random $\eta$ and $\phi$ directions.}
\label{fig:phys_validation}
\end{figure}

In order to test the computing performance of \GEANTV in \CMSSW and compare it to \GEANTfour,
it must first be established that the two simulation engines perform similar operations when given the same input.
This is demonstrated by running both detector simulation packages with the exact same generated events:
single electrons with $E=50\unit{GeV}$ and random $\eta$ and $\phi$ directions.
The constant magnetic field is enabled and multiple threads are used to process the events.
Figure~\ref{fig:phys_validation} shows the results in the ECAL barrel,
which confirm that the energy, time, and number of hits agree within 5\% between \GEANTfour and \GEANTV.

The computing performance tests use the latest version of \GEANTV~\cite{GeantVSW} with vectorized multiple scattering and magnetic field propagation enabled.
The generated sample comprises 500 events with two electrons, each having $E=50\unit{GeV}$ and random $\eta$ and $\phi$ directions.
These generated events are copied and concatenated multiple times for the multithreaded tests, so that each thread processes the same events on average.
Accordingly, the number of events processed per thread is kept constant in the tests,
and any unused threads are kept busy to prevent any changes in allowed clock speed based on the CPU load.
The \ROOT file output is disabled, as those operations are not thread-safe and therefore decrease the performance artificially.
The Run 2 geometry of the CMS detector is used in the simulation, with the adapted calorimeter scoring code enabled.
The dedicated test machine has an Intel{\textregistered} Xeon{\textregistered} CPU E5-2683 v3 with a 2.00\unit{GHz} clock speed,
35840\unit{KB} cache, 28 cores, and support for sse4.2 instructions.

\begin{figure}[h]
\centering
\includegraphics[width=0.49\linewidth]{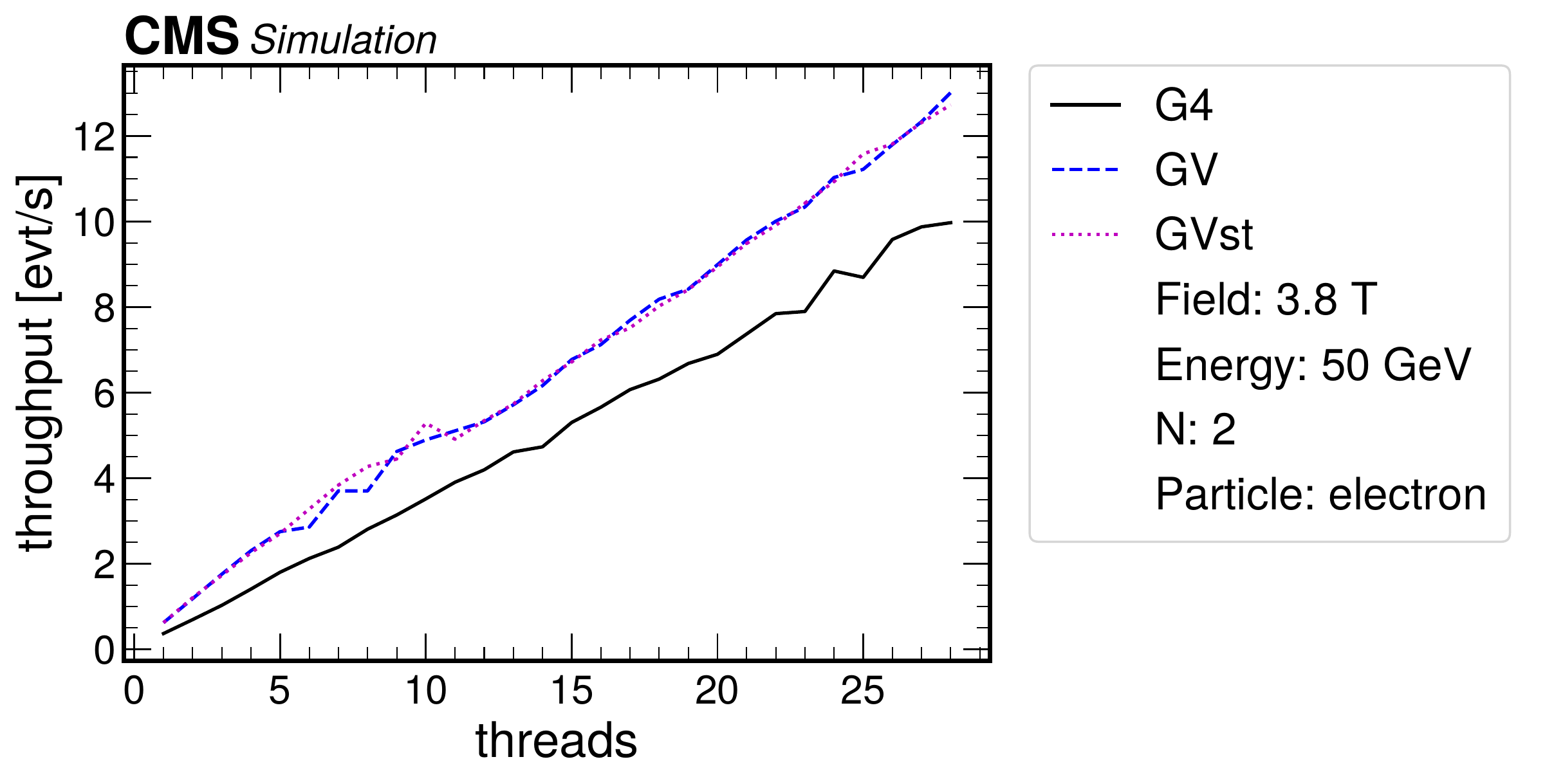}
\includegraphics[width=0.49\linewidth]{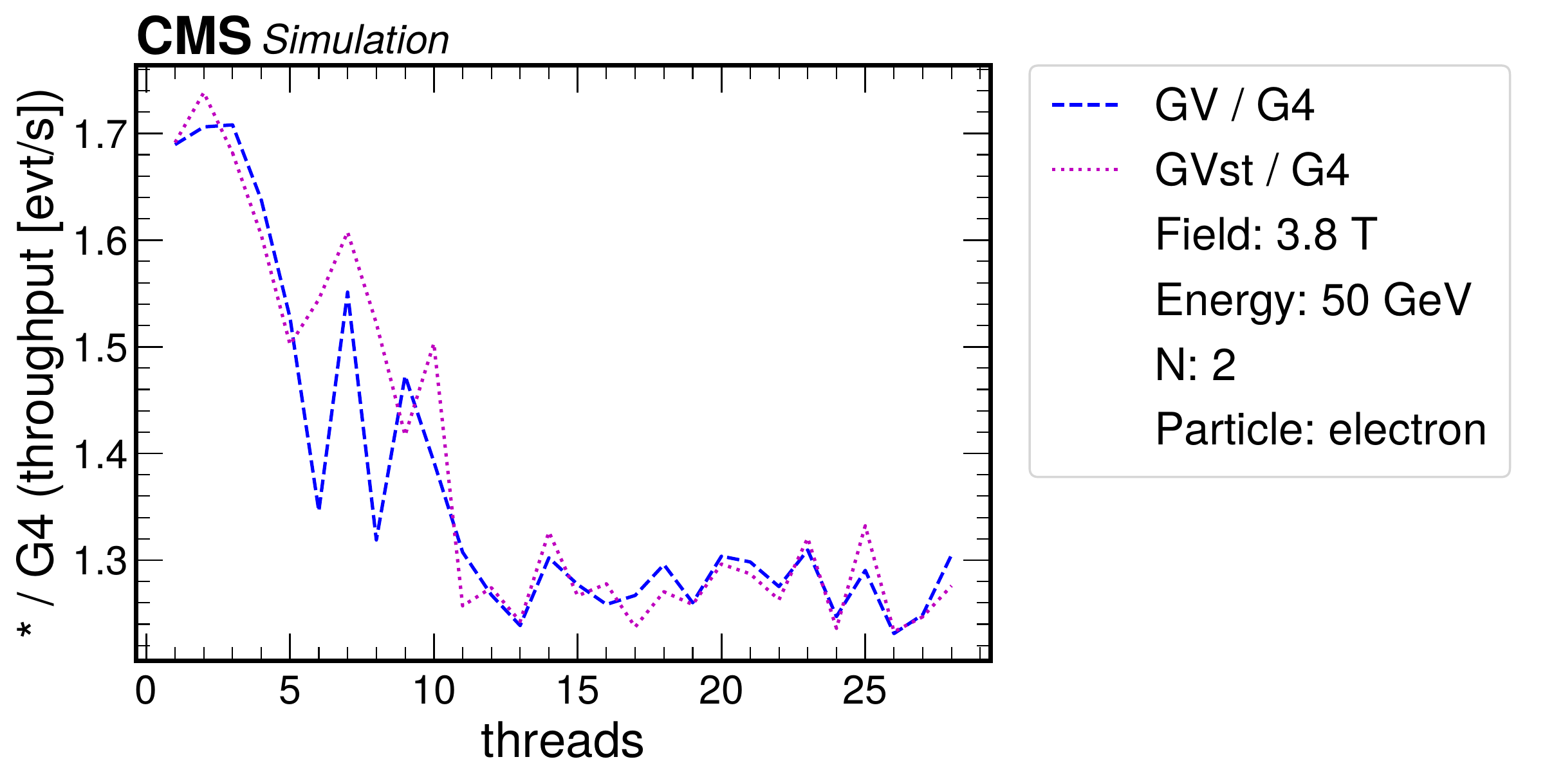}\\
\includegraphics[width=0.49\linewidth]{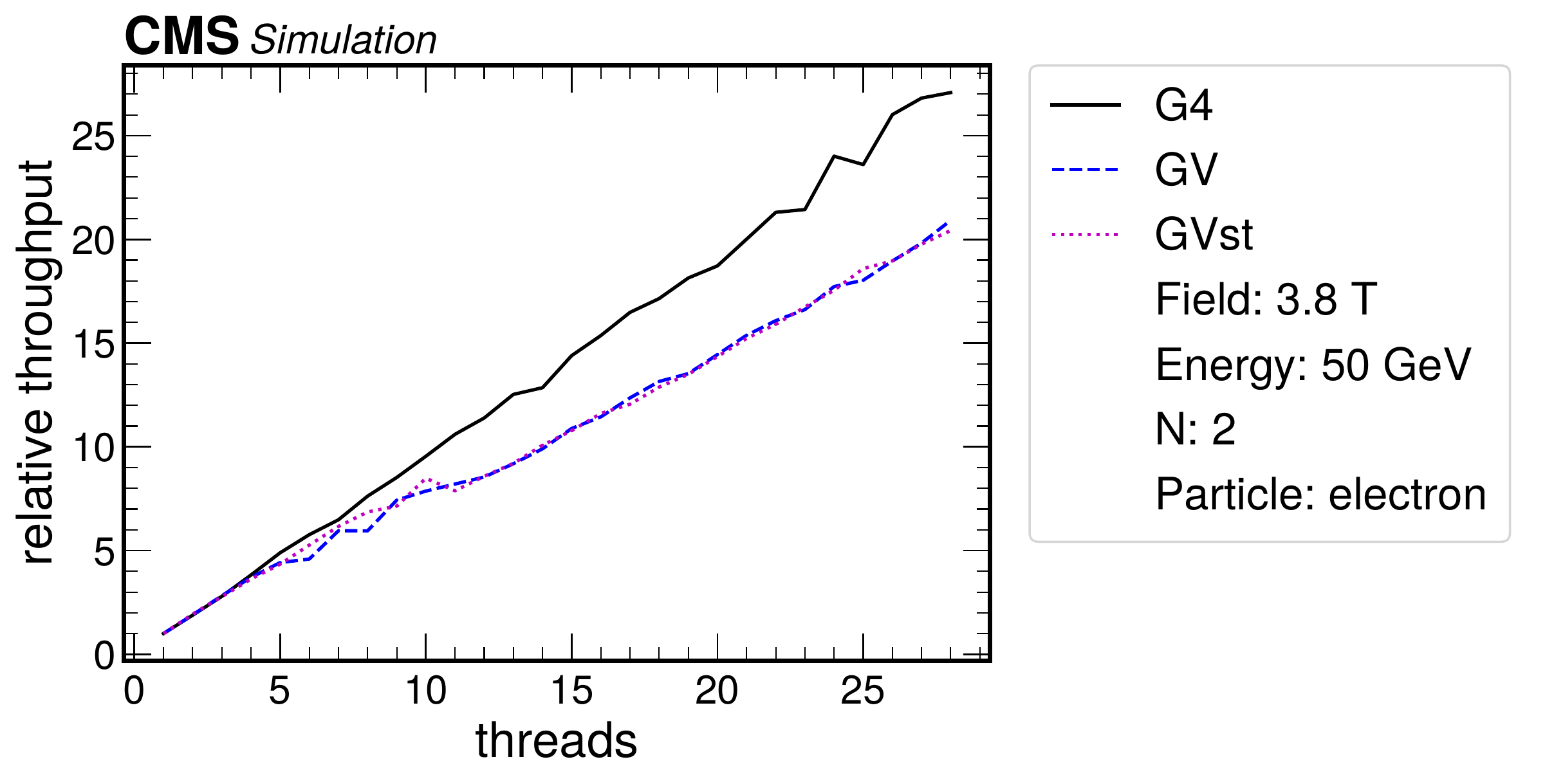}
\caption{Top left: throughput in events per second for \GEANTfour and \GEANTV, versus the number of threads.
Top right: the ratio of \GEANTV to \GEANTfour throughput, versus the number of threads.
Bottom: The relative throughput for multiple threads compared to a single thread for \GEANTfour and \GEANTV.
In all plots, the item ``GVst'' indicates \GEANTV in single-track mode, as explained in the text.}
\label{fig:time_perf}
\end{figure}

\begin{figure}[h]
\centering
\includegraphics[width=0.49\linewidth]{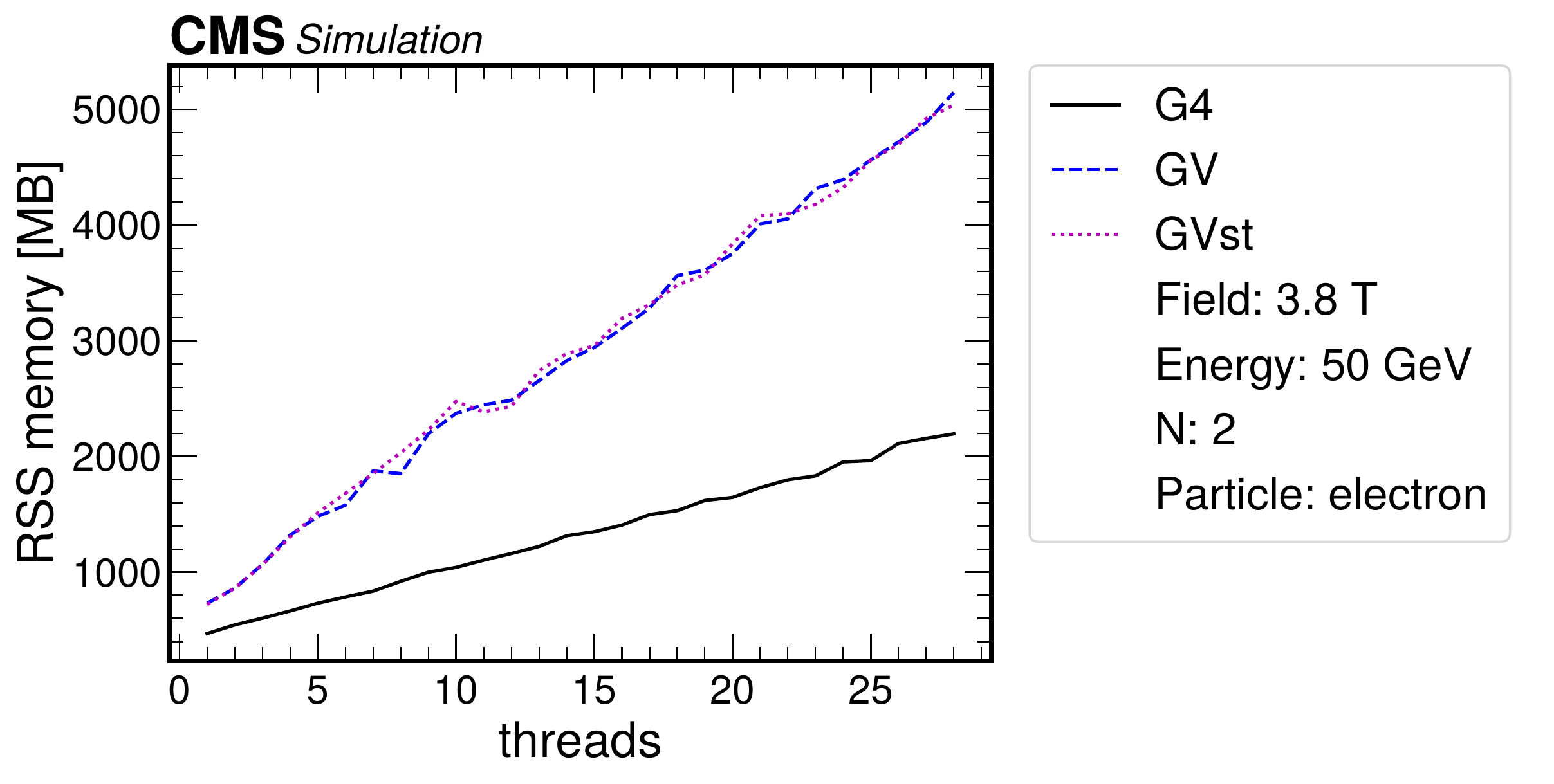}
\includegraphics[width=0.49\linewidth]{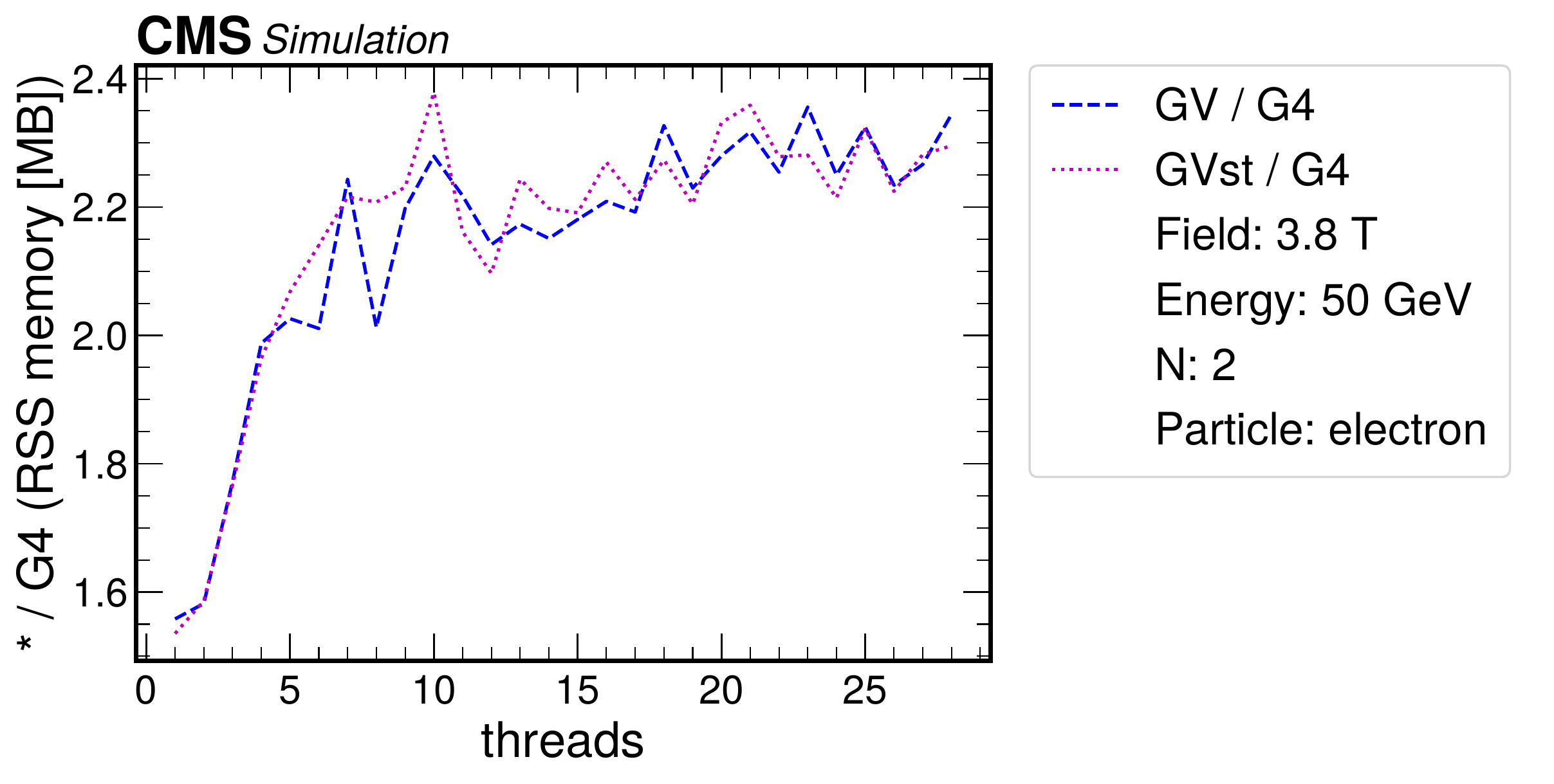}
\caption{Left: RSS memory in MB for \GEANTfour and \GEANTV, versus the number of threads.
Right: the ratio of \GEANTV to \GEANTfour RSS memory, versus the number of threads.
In all plots, the item ``GVst'' indicates \GEANTV in single-track mode, as explained in the text.}
\label{fig:mem_perf}
\end{figure}

To characterize the performance results in the full \CMSSW framework,
the standalone \GEANTV tests with the CMS geometry are run using a single thread,
with the same settings as above and on the same dedicated machine.
These tests show that the standalone \GEANTV is 1.6 times faster than \GEANTfour.
The \CMSSW test actually achieves a larger improvement factor of 1.7,
as shown in Fig.~\ref{fig:time_perf} using the event throughput calculated from the wall time measurement.
However, when multiple threads are used, the improvement factor drops to 1.3,
indicating that \GEANTV does not scale as well as \GEANTfour.
This can also be seen in Fig.~\ref{fig:time_perf}
from the relative throughput factor computed as the ratio of the throughput for $N$ threads to the throughput for 1 thread.
These tests also compare an alternative \GEANTV mode in which tracks are not basketized,
but instead processed serially just as in \GEANTfour.
The performance of this single track mode is equivalent to the basketized mode,
indicating that the gains from vectorization are negated by the overhead from basketization.
Figure~\ref{fig:mem_perf} shows the RSS memory usage, which grows linearly with the number of threads, as expected.
\GEANTV uses more memory than \GEANTfour, also as expected; it would still be able to run efficiently on grid computing nodes.

\section{Conclusions}\label{sec:conclude}

The CMS detector simulation for the HL-LHC upgrades may be 2--3 times slower than the Run 2 simulation.
This increase in CPU usage is primarily driven by the High Granularity Calorimeter upgrade,
which has a more complex geometry and requires more precise physics models.
Ongoing efforts to improve the CPU performance the simulation have been fruitful,
but more involved research and development is needed to meet the challenges posed by the HL-LHC.

The effort to integrate the \GEANTV vectorized transport engine in the CMS software successfully addressed all goals.
Co-development between the different developers ensured compatible threading models and interfaces.
A similar, and even slightly larger, improvement factor was measured using the full experiment software framework, compared to standalone tests.
This strong performance arises from the underlying improvement in \GEANTV,
which is reduced instruction cache misses from a smaller compiled library.
Because the CMS framework loads additional libraries compared to standalone execution of \GEANTV, the smaller library size is even more impactful.
Finally, an efficient path to integration has been established, in order to minimize personpower needs
for testing of \GEANTV or any future simulation R\&D products in experiment software frameworks.

\bibliography{cms}

\begin{thebibliography}{25}

\bibitem{CMS-TDR-019}
{CMS Collaboration}, CMS Technical Design Report CERN-LHCC-2017-023,
  CMS-TDR-019, CERN (2017),
  \urlstyle{tt}\url{https://cds.cern.ch/record/2293646}

\bibitem{Apostolakis:2018ieg}
J.~Apostolakis et~al. (HEP Software Foundation) (2018), \texttt{1803.04165}

\bibitem{Geant4SW}
{Geant4 project}, \emph{Geant4}, [software] (2016), version 10.02.p02 (accessed
  2018-09-04),
  \urlstyle{tt}\url{https://github.com/Geant4/geant4/releases/tag/v10.2.2}

\bibitem{Agostinelli:2002hh}
S.~Agostinelli et~al., Nucl. Instrum. Meth. A \textbf{506}, 250 (2003)

\bibitem{Allison:2016lfl}
J.~Allison et~al., Nucl. Instrum. Meth. A \textbf{835}, 186 (2016)

\bibitem{Alves:2017she}
J.~Albrecht et~al. (HEP Software Foundation), Comput. Softw. Big Sci.
  \textbf{3}, 7 (2019), \texttt{1712.06982}

\bibitem{Chatrchyan:2008zzk}
{CMS Collaboration}, JINST \textbf{3}, S08004 (2008)

\bibitem{CMS-TDR-010}
{CMS Collaboration}, CMS Technical Design Report CERN-LHCC-2012-015,
  CMS-TDR-010, CERN (2012),
  \urlstyle{tt}\url{https://cds.cern.ch/record/1481837}

\bibitem{KADRI20093624}
O.~Kadri, V.~Ivanchenko, F.~Gharbi, A.~Trabelsi, Nucl. Instrum. Meth. B
  \textbf{267}, 3624 (2009)

\bibitem{CMSSW}
{CMS Collaboration}, \emph{{CMSSW}}, [software] (2019), (accessed 2019-10-01),
  \urlstyle{tt}\url{https://github.com/cms-sw/cmssw}

\bibitem{Pedro:2019mkq}
K.~Pedro (CMS), Eur. Phys. J Web Conf. \textbf{214}, 02036 (2019)

\bibitem{Lange:2015sba}
D.J. Lange, M.~Hildreth, V.N. Ivantchenko, I.~Osborne (CMS), J. Phys. Conf.
  Ser. \textbf{608}, 012056 (2015)

\bibitem{VecGeomSW}
{VecGeom project}, \emph{{VecGeom}}, [software] (2019), version 01.01.01
  (accessed 2019-10-01),
  \urlstyle{tt}\url{https://gitlab.cern.ch/VecGeom/VecGeom/tags/v00.05.00}

\bibitem{Apostolakis:2015raa}
J.~Apostolakis et~al., J. Phys. Conf. Ser. \textbf{608}, 012023 (2015)

\bibitem{VecCoreSW}
{VecCore project}, \emph{{VecCore}}, [software] (2018), version 0.5.2 (accessed
  2019-10-01),
  \urlstyle{tt}\url{https://github.com/root-project/veccore/releases/tag/v0.5.2}

\bibitem{VecMathSW}
{VecMath project}, \emph{{VecMath}}, [software] (2018), version 077152d
  (accessed 2019-10-01),
  \urlstyle{tt}\url{https://github.com/root-project/vecmath}

\bibitem{amadio2020geantv}
G.~Amadio, A.~Ananya, J.~Apostolakis, M.~Bandieramonte, S.~Banerjee,
  A.~Bhattacharyya, C.~Bianchini, G.~Bitzes, P.~Canal, F.~Carminati et~al.,
  \emph{{GeantV}: {Results} from the prototype of concurrent vector particle
  transport simulation in {HEP}} (2020), submitted to \emph{Comp. Soft. Big
  Sci.}, \texttt{2005.00949}

\bibitem{SimGVCoreSW}
K.~Pedro, \emph{{SimGVCore v1.0.0}}, [software] (2019), (accessed 2019-11-05),
  \urlstyle{tt}\url{https://github.com/kpedro88/SimGVCore/tree/v1.0.0}

\bibitem{HepMC2SW}
{HepMC project}, \emph{{HepMC2}}, [software] (2012), version 2.06.07 (accessed
  2018-09-04),
  \urlstyle{tt}\url{http://lcgapp.cern.ch/project/simu/HepMC/download/HepMC-2.06.07.tar.gz}

\bibitem{Dobbs:2001ck}
M.~Dobbs, J.B. Hansen, Comput. Phys. Commun. \textbf{134}, 41 (2001)

\bibitem{Osborne_2014}
I.~Osborne, E.~Brownson, G.~Eulisse, C.D. Jones, D.J. Lange, E.~Sexton-Kennedy,
  J. Phys. Conf. Ser. \textbf{513}, 022026 (2014)

\bibitem{RootSW}
{ROOT project}, \emph{{ROOT}}, [software] (2018), version 6.12.07 (accessed
  2019-10-01),
  \urlstyle{tt}\url{https://github.com/root-project/root/releases/tag/v6-12-06}

\bibitem{Brun:1997pa}
R.~Brun, F.~Rademakers, Nucl. Instrum. Meth. A \textbf{389}, 81 (1997)

\bibitem{makortelCHEP2019}
A.~Bocci, D.~Dagenhart, V.~Innocente, M.~Kortelainen, F.~Pantaleo, M.~Rovere,
  \emph{Bringing heterogeneity to the {CMS} software framework},
  arXiv:2004.04334 (2019), submitted to \textit{Eur. Phys. J. Web Conf.}
  (CHEP2019), \texttt{2004.04334}

\bibitem{GeantVSW}
{GeantV project}, \emph{{GeantV}}, [software] (2019), version 63468c9b
  (pre-beta-7+) (accessed 2019-10-01),
  \urlstyle{tt}\url{https://gitlab.cern.ch/GeantV/geant}

\end{thebibliography}

\end{document}